%Paper: hep-th/9410127
%From: "Don Spector" <das@hepth.cornell.edu>
%Date: Tue, 18 Oct 1994 13:40:08 -0400

%
\input phyzzx.tex
\hsize=6truein
\vsize=8.5truein
\tenrm\baselineskip=14pt
%\singlespace

\def\B{Bogomol'nyi }

\def\Stilde{{\tilde S}}
\def\semi{;\hfill\break}

\REF\tnc{A.M. Polyakov, {\it JETP Lett.} {\bf 20} (1974) 194;
A.A. Belavin, A.M.
Polyakov, A.S. Schwartz, and Yu.S. Tyupkin, {\it Phys. Lett.}
{\bf 59B} (1975) 85;
G. 't Hooft, {\it Phys. Rev.} {\bf D14} (1976) 3432.}
\REF\bounds{M.K. Prasad and C.H.
Sommerfield, {\it Phys. Rev. Lett.} {\bf 35} (1975) 760;
E.B. \B , {\it Sov. J. Nucl. Phys.}
{\bf 24} (1976) 449;
G. 't Hooft, {\it Phys. Rev.
Lett.} {\bf 37} (1976) 8;
A.M. Polyakov, {\it Nucl. Phys.} {\bf B121} (1977) 429.}
\REF\ntwo{D. Olive and E. Witten,{\it Phys.
Lett.} {\bf 78B} (1978) 97;
C. Lee, K. Lee, and E. Weinberg,
{\it Phys. Lett.} {\bf 243B} (1990) 105;
J.A. de Azcarraga, J.P. Gauntlett,
J.M. Izquierdo, and P.K. Townsend, {\it Phys. Rev. Lett.}
{\bf 63} (1989) 2443; M. Cvetic, F. Quevedo, and S.-J. Rey,
{\it Phys. Rev. Lett.} {\bf 67} (1991) 1836;
J. Edelstein, C. Nunez, and F. Schaposnik, {\it Phys.Lett.}
{\bf B329} (1994) 39.}
\REF\us{Z. Hlousek and D. Spector, {\it Nucl. Phys.} {\bf
B370} (1992) 143;
Z. Hlousek and D. Spector, {\it Phys. Lett.}
{\bf 283B} (1992) 75;
Z. Hlousek and D. Spector,
{\it Mod. Phys. Lett.} {\bf A7} (1992) 3403;
Z. Hlousek and D. Spector,
{\it Nucl. Phys.} {\bf B397} (1993) 173;
Z. Hlousek and D. Spector, {\it Solitons and Instantons without
Supersymmetry}, submitted for publication.}
\REF\pot{C. Cronstr\"om and J.
Mickelsson, {\it J. Math. Phys.} {\bf 24} (1983) 2528;
R. Jackiw {\it in} S. Treiman, R. Jackiw, B. Zumino, and E.
Witten {\it Current Algebra and Anomalies} (World Scientific,
Singapore, 1985).}
\REF\hls{R. Haag, J. {\L }opusza\'nski, and M. Sohnius, {\it Nucl. Phys.}
{\bf B288} (1975) 257.}
\REF\daddiv{ A. D'Adda and P. Di Vecchia, {\it Phys.Lett.}
{\bf 73B} (1978) 162.}

\vglue -.5in
\centerline{Sept. 1994 \hfill HWS-94/16}
\vglue .5cm
\centerline{\bf Supersymmetry and Solitons: $N=2$ and $N=0$}
\vglue .7cm
\tenrm\baselineskip=12pt
\centerline{DONALD SPECTOR}\foot{spector@hws.edu}
\centerline{\tensl Dept. of Physics, Hobart and William Smith
Colleges, Geneva, NY 14456 USA}
\vglue .3cm
\centerline{and}
\vglue .3cm
\centerline{ZVONIMIR HLOUSEK}\foot{hlousek@beach.csulb.edu}
\centerline{\tensl Dept. of Physics and Astronomy, California State
U., Long Beach,
CA 90840 USA}
\vglue .7cm
\centerline{ABSTRACT}
{\rightskip=3pc
 \leftskip=3pc
 \tenrm\baselineskip=12pt\parindent=0pc
This talk summarizes our recent work establishing an algebraic,
model-independent basis for the existence of \B bounds and
\B equations for topologically non-trivial solitons and instantons.
Our arguments use supersymmetry in an essential way to understand
both supersymmetric and non-supersymmetric theories.  Our
arguments are constructive and work in nearly any number of dimensions.
\vglue  .1cm}
\centerline{\tensl Presented at and to appear in the proceedings of the
XX${}^{th}$ International Colloquium on}
\centerline{\tensl Group Theoretical Methods in Physics, Osaka, Japan,
July 1994}

\twelverm\baselineskip=14pt

\chapter{\twelvebf Introduction and Overview}

Theories with topologically
non-trivial solitons and instantons$^\tnc$ regularly
exhibit \B or self-duality bounds (the
soliton energy or instanton action is bounded from below
by the charge or instanton number, respectively), with field
configurations that saturate such bounds satisfying
first-order differential equations, called \B or
self-duality equations.$^\bounds$  (We will refer to such bounds
and equations collectively as {\it \B relationships}.)
Despite this regularity, these features have so far
been understood only on a case-by-case basis, by invoking
the equations of motion.

Similarly, in models with $N=1$ supersymmetry and a conserved
topological charge, one regularly
finds a larger algebraic symmetry structure, namely $N=2$ supersymmetry
with the topological charge as a central charge,$^\ntwo$
and yet this, too, has heretofore been understood only
on a case-by-case basis.

We address these theoretical shortcomings here by establishing
the above results for solitons and instantons\foot{We will use the
terms {\it soliton} and {\it instanton} to refer to any field configuration
with non-trivial topological charge or instanton number.  This liberal
usage will be useful here.}
in a general, model-independent way.
Our method is to obtain these results first in the supersymmetric case,
and then to show that this implies the corresponding results in the
generic non-supersymmetric case.
Our use of supersymmetry to
understand non-supersymmetric theories is technically very similar
to the use of complex analysis to address question about real functions.

Because of space limitations, we will be very brief here.
The reader interested in the subtleties and technicalities should
refer to our papers for a detailed accounting.$^\us$

\chapter{\twelvebf Why Topological Charges
Imply Extended Supersymmetry}

We demonstrate here that a theory with an $N=1$ supersymmetry and
a conserved topological charge necessarily has an $N=2$ supersymmetry
in which the topological charge appears as the central charge.
We first give the argument in $2+1$ dimensions, then generalize this
to arbitrary higher dimension, and finally discuss the
significance of the breakdown of our argument in $1+1$ dimensions.

A theory with $N=1$ supersymmetry and a topologically conserved charge
in $2+1$ dimensions has a conserved real spinor charge $Q^\alpha$, with
{\vskip -.5cm}
$$\{Q^\alpha,Q^\beta\} = P^{\alpha\beta}~~,\eqno(1)$$
and a current $J_\mu$ for which $\partial^\mu J_\mu= 0 $ can be
derived without using the equations of motion, which means that
one can write
$J_\mu$ in terms of a vector potential $A_\nu$ via$^\pot$
{\vskip -.5cm}
$$J_\mu = \epsilon_{\mu\nu\lambda}\partial^\nu A^\lambda~~.\eqno(2)$$
The gauge
equivalent vector potentials $A_\mu$ and $A_\mu + \partial_\mu \chi$
produce the same topological current $J_\mu$.  Among the gauge equivalent
potentials, there is one that is divergenceless; hereon, we use
$A_\mu$ to refer to this particular potential.  (Note that the theory
has no gauge {\it symmetry}, so we are not gauge-fixing
the theory.) Define
{\vskip -.5cm}
$${\tilde S}^\alpha_\mu = [Q^\alpha,A_\mu]~~.\eqno(3)$$
Since supersymmetry transformations commute with translations, we have
$\partial^\mu {\tilde S}^\alpha_\mu = 0$, i.e., ${\tilde S}^\alpha_\mu$
is a conserved vector-spinor.

Under the original supersymmetry, ${\tilde S}^\alpha_\mu$
transforms into the non-trivial conserved topological charge, so this
conserved spinor current is neither trivial nor is it the
the original supercurrent.
Consequently, ${\tilde S}^\alpha_\mu$ must be a second
conserved spinor current.
Since under the original supersymmetry this
new supercurrent transforms into the topological current, the
theory is invariant under
an $N=2$ superalgebra with a central charge given by
the topological charge.

Had we started from a
different but gauge-equivalent potential, the corresponding vector-spinor
produced would differ from the second supercurrent by an element of the
kernel of the original supercharge.  Modding out by this kernel gives
another means of identifying the physical supercurrent from among these
vector-spinors.

In higher dimensions, one can write a topologically conserved
current as the curl of a
divergenceless $d-2$-index antisymmetric
tensor (where $d$ is the spacetime dimension).
The second
supercurrent is then simply
{\vskip -.5cm}
$${\Stilde}^\alpha_{\mu_1} =
  [Q^\alpha,A_{\mu_1 \mu_2 \cdots \mu_{d-2}}]\gamma^{\mu_2}\cdots
   \gamma^{\mu_{d-2}}~~.\eqno(4)$$
The rest of the argument proceeds as before.

Note that our construction explicitly breaks down in $1+1$ dimensions.
This is as it must be.
In $1+1$ dimensions, there are supersymmetric models in which the
topological charge does not serve as a central charge.
That our method handles arbitrary
dimensions, but explicitly breaks down for $1+1$ dimensions,
indicates that we have indeed identified a fundamental approach to this
phenomenon.

\chapter{\twelvebf \B Relationships for
Supersymmetric Soliton Theories}

The extended superalgebra derived above implies that \B relationships
arise in any supersymmetric theory with a topological charge.
This result is easy to derive, using a standard observation.  For
simplicity, we give only the $2+1$ dimensional case here.

 From the above results, we see we have the algebraic relation
{\vskip -.5cm}
$$\{Q^\alpha_L , Q^\beta_M\} = P^{\alpha\beta}\delta_{LM}
      + T \epsilon^{\alpha\beta}\epsilon_{LM}~~,\eqno(5)$$
where $Q^\alpha_L$, $L=1,2$, are the two supercharges,
$P^{\alpha\beta}$ is the
momentum, and $T$ is the topological charge.
Since this anticommutator is hermitian, its square is positive
semi-definite; taking the trace of this square
we obtain the \B bound
{\vskip -.5cm}
$$M^2 - T^2 \ge 0~~,\eqno(6)$$
where $M$ is the rest mass.
Furthermore, this bound is saturated only when
the field configuration is annihilated by one of the supercharges, a
condition represented by a set of first-order equations (since the
supercharges can be represented by first-order differential operators).
These, then, are the \B equations of the theory.

\chapter{\twelvebf \B Relationships for Solitons in General}

Consider a generic, non-supersymmetric Lagrangian ${\cal L}$
which is a functional of some field(s) $\phi$, and which
has a conserved topological charge $T[\phi]$. The energy of a field
configuration is given by a functional $E[\phi]$.  We now demonstrate
that this theory exhibits \B relationships.

Now consider a supersymmetric extension of this theory.
Such an extension has a Lagrangian ${\cal L}_s$ which is a functional
of the original field(s) $\phi$ and some additional field(s) $\psi$.
(This notation is suggestive of scalar and fermionic fields, but in
fact any theory can be suitably extended.$^\us$)
This theory has a topological
charge $T_s[\phi,\psi]$ and an energy functional $E_s[\phi,\psi]$.
There are three important features of this extension:

1. The field configurations of the original theory are also field
configurations of the extended theory.

2. Since the topological charge is conserved without reference to
the equations of motion, the extended theory has the {\it exact
same} topological charge as the original theory, and so
$T_s[\phi,\psi] = T_0[\phi]$, irrespective of the value of $\psi$.

3. We can and do choose the extension such
that  $E_s[\phi,\psi=0] = E_0[\phi]$.

The extended theory is a supersymmetric theory with a conserved
topological charge.  Thus,
$E_s[\phi,\psi] \ge |T_s[\phi,\psi] |$ for any field configuration,
and any field configuration that saturates this inequality
satisfies first-order \B equations.
For field configurations for which
$\psi=0$, using points 2 and 3 above, the preceding inequality yields
{\vskip -.5cm}
$$E_0[\phi] \ge |T_0[\phi]|~~.\eqno(7)$$
In this way, we have just obtained the \B bound of the original theory!
By points 2 and 3 again, a field configuration that saturates
the \B bound
of the original theory also saturates the
supersymmetric \B bound, and so this field configuration satisfies
first-order differential equations when viewed as a configuration
of the supersymmetric extension.  These equations involve
only $\phi$ (since $\psi=0$), and so they are the sought-after
\B equations of the original theory, derived in a general way.

\chapter{\twelvebf Instantons and Self-Duality in General}

To extend these results to topologically non-trivial instantons is the
obvious next step.
Since instanton number is {\it not} a
conserved charge, it cannot
be woven directly into a superalgebra.
Fortunately, we can still
build on our previous results.

Consider a Euclidean field theory
in $d$ dimensions with action $S_d[\phi]$ which has a topological
instanton number given by the functional $I_d[\phi]$.  One can construct
an associated $d+1$-dimensional Minkowskian theory, by making all fields
functions of an additional time coordinate, and adding the necessary
field components (e.g., temporal components to vector fields)
and time derivatives
to ensure Lorentz invariance.  The Minkowskian theory has
energy functional $E_{d+1}$.  Furthermore, the instanton number of the
Euclidean theory becomes a topologically conserved charge $T_{d+1}$ in the
Minkowskian theory,
since the $d$-dimensional instanton number cannot change as the fields
evolve continuously from one
time-slice to the next in $d+1$ dimensions.

The $d+1$ dimensional theory exhibits \B relationships, as we have argued
above.  But the instantons of the $d$-dimensional theory are static
solitons of the Minkowskian theory, which thus
obey $E_{d+1}\ge |T_{d+1}|$.  When these field configurations are viewed
as instantons, this inequality becomes $S_d\ge |I_d|$.  Furthermore,
if an instanton saturates this bound, it satisfies a \B equation in $d+1$
dimensions when viewed
as a soliton.  Because
the field is static, this in fact is a first-order equation in $d$ dimensions,
that is, the Euclidean space instanton
self-duality equation, now obtained in a perfectly general
way, and in an essentially unified treatment with solitons.

\chapter{\twelvebf Closing Remarks}

Our methods have demonstrated in a model-independent
way that \B relationships must arise for
topological solitons and instantons.
It is the hidden hand of supersymmetry that
makes such a general approach feasible.
Our method offers other insights, as well.  Our approach shows very
naturally why zero modes in self-dual or Bogomol'nyi-saturating
backgrounds must be associated with index theorems.  We are also able
to generalize to any model a result of D'Adda and Di Vecchia's$^\daddiv$
that in certain theories the non-zero modes in a Bogomol'nyi-saturating
soliton or instanton background are bose--fermi degenerate.
Further, there are suggestive connections between our results
and the construction of topological field theories from $N=2$
supersymmetric ones.

\ack The research of D.S. was supported in part by  NSF Grant
No. PHY-9207859.
%\refout
%\vglue 1cm

\centerline{\twelvebf References}
\tenrm\baselineskip=12pt
\begingroup \refoutspecials \catcode `\^^M=10
\refitem {1}
A.M. Polyakov, {\tensl JETP Lett.} {\tenbf 20} (1974) 194\semi
A.A. Belavin, A.M.
Polyakov, A.S. Schwartz, and Yu.S. Tyupkin, {\tensl Phys. Lett.}
{\tenbf 59B} (1975) 85\semi
G. 't Hooft, {\tensl Phys. Rev.} {\tenbf D14} (1976) 3432.
\refitem {2}
M.K. Prasad and C.H.
Sommerfield, {\tensl Phys. Rev. Lett.} {\tenbf 35} (1975) 760\semi
E.B. Bogomol'nyi, {\tensl Sov. J. Nucl. Phys.}
{\tenbf 24} (1976) 449\semi
G. 't Hooft, {\tensl Phys. Rev.
Lett.} {\tenbf 37} (1976) 8\semi
A.M. Polyakov, {\tensl Nucl. Phys.} {\tenbf B121} (1977) 429.
\refitem {3}
D. Olive and E. Witten,{\it Phys.
Lett.} {\tenbf 78B} (1978) 97\semi
J.A. de Azcarraga, J.P. Gauntlett,
J.M. Izquierdo, and P.K. Townsend, {\tensl Phys. Rev. Lett.}
{\tenbf 63} (1989) 2443\semi
C. Lee, K. Lee, and E. Weinberg,
{\tensl Phys. Lett.} {\tenbf 243B} (1990) 105\semi
 M. Cvetic, F. Quevedo, and S.-J. Rey,
{\tensl Phys. Rev. Lett.} {\tenbf 67} (1991) 1836\semi
J. Edelstein, C. Nunez, and F. Schaposnik, {\tensl Phys.Lett.}
{\tenbf B329} (1994) 39.
\refitem {4}
Z. Hlousek and D. Spector, {\tensl Nucl. Phys.} {\tenbf
B370} (1992) 143\semi
Z. Hlousek and D. Spector, {\tensl Phys. Lett.}
{\tenbf 283B} (1992) 75\semi
Z. Hlousek and D. Spector,
{\tensl Mod. Phys. Lett.} {\tenbf A7} (1992) 3403\semi
Z. Hlousek and D. Spector,
{\tensl Nucl. Phys.} {\tenbf B397} (1993) 173\semi
Z. Hlousek and D. Spector, {\tensl Solitons and Instantons without
Supersymmetry}, submitted for publication.
\refitem {5}
C. Cronstr\"om and J.
Mickelsson, {\tensl J. Math. Phys.} {\tenbf 24} (1983) 2528\semi
R. Jackiw {\tensl in} S. Treiman, R. Jackiw, B. Zumino, and E.
Witten {\tensl Current Algebra and Anomalies} (World Scientific,
Singapore, 1985).
\refitem {6}
R. Haag, J. {\L }opusza\'nski, and M. Sohnius, {\tensl Nucl. Phys.}
{\tenbf B288} (1975) 257.
\refitem {7}
 A. D'Adda and P. Di Vecchia, {\tensl Phys.Lett.}
{\tenbf 73B} (1978) 162.
\par \endgroup

\bye